\documentclass[journal=jceaax,manuscript=article]{achemso}

\usepackage{chemformula}
\usepackage{bm}
\usepackage{natbib}
\usepackage[T1]{fontenc}
\usepackage[utf8]{inputenc}
\usepackage{subcaption}
\usepackage{amssymb}
\usepackage{multirow}
\usepackage{longtable}
\usepackage{rotating}
\usepackage{hhline}
\usepackage{amsmath}
\usepackage{graphicx}
\usepackage{color}
\usepackage{soul}
\usepackage{xcolor}
\usepackage{tikz}
\usepackage{blindtext}
\usepackage{siunitx}
\DeclareSIUnit[quantity-product = ]\percent{\char`\%}
\usetikzlibrary{shapes.geometric}
\usepackage{booktabs}
\usepackage{tabularx}
\usepackage{caption}

\author{Dominik~Gond}
\affiliation{Laboratory of Engineering Thermodynamics, RPTU Kaiserslautern, 67663 Kaiserslautern, Germany}

\author{Justus~Arweiler}
\affiliation{Laboratory of Engineering Thermodynamics, RPTU Kaiserslautern, 67663 Kaiserslautern, Germany}

\author{Thomas~Specht}
\affiliation{Laboratory of Engineering Thermodynamics, RPTU Kaiserslautern, 67663 Kaiserslautern, Germany}

\author{Hans~Hasse}
\affiliation{Laboratory of Engineering Thermodynamics, RPTU Kaiserslautern, 67663 Kaiserslautern, Germany}

\author{Fabian~Jirasek\textsuperscript}
\affiliation{Laboratory of Engineering Thermodynamics, RPTU Kaiserslautern, 67663 Kaiserslautern, Germany}
\email{fabian.jirasek@rptu.de}

\title{CHAOS - A Large-scale Database for $\sigma$-Profiles and Other Molecular Descriptors}

\begin{document}

\begin{abstract}
$\sigma$-profiles obtained from quantum-chemical calculations are key molecular descriptors for solvent selection, thermodynamic modeling, and data-driven molecular design. However, existing $\sigma$-profile libraries are limited in size and are difficult to combine into a single consistent dataset because they were generated using different quantum-chemical methods. In this work, we introduce CHAOS (Computed High-Accuracy Observables and Sigma Profiles), a large-scale and internally consistent database providing $\sigma$-profiles for 53\,091 molecules, along with additional quantum-chemical observables including gas-phase geometries, single-point conductor-like polarizable continuum (C-PCM) data, infrared spectra, ideal-gas heat capacities and entropies, and atomic orbital nuclear magnetic resonance (NMR) shielding tensors. All data were generated using a standardized quantum-chemical workflow based on an $\omega$B97X-D/def2-TZVP level of theory. The CHAOS database covers molecules composed of a diverse set of elements, with molecular masses up to 400\,amu and dipole moments up to 15\,D, and is freely available on Zenodo under an open license. It extends the number of molecules for which $\sigma$-profiles are publicly available by more than an order of magnitude and systematically links them to a broad range of other quantum-chemical molecular descriptors. CHAOS provides a comprehensive and consistent foundation for developing models of molecular and thermodynamic properties -- both physics-based and machine-learning approaches -- across chemistry, chemical engineering, and materials science, greatly extending the possibilities and the available quantum-chemical data basis. 
\end{abstract}

\section{Introduction}
\label{sec:intro}

The $\sigma$-profile is a molecular descriptor, derived within the COSMO (Conductor-like Screening Model) framework \cite{klamt1995conductor} from quantum-chemical calculations. It represents the molecular surface charge density distribution and thereby encodes information on polarity and hydrogen-bonding characteristics, which form the basis for modeling intermolecular interactions, including dispersion, in COSMO-based methods. It is the central descriptor of COSMO-RS (Conductor-like Screening Model for Real Solvents) \cite{klamt2005cosmo}, an established model for the Gibbs excess energy that enables the prediction of activity coefficients in any mixture for which the $\sigma$-profiles of all constituent components are available.

$\sigma$-profiles have been applied in several thermodynamic models. Besides the most prominent COSMO-RS by Klamt and Eckert \cite{klamt2005cosmo}, other variants include the COSMO-SAC model by Lin and Sandler \cite{lin2002priori} and its later reparameterizations \cite{hsieh2010improvements, paulechka2015reparameterization}, the COSMO-RS(Ol) version developed by Mu and Gmehling \cite{mu2007group}, and hybrid equation-of-state approaches, such as the PR+COSMO-SAC model by Tsai and co-workers \cite{hsieh2012first, tsai2023improved}. Beyond the COSMO-RS family, $\sigma$-profiles have also been incorporated into hybrid group-contribution models, such as COSMO-UNIFAC \cite{dong2018united, dong2020cosmo}, cubic-equation extensions like COSMO-SRK and COSMO-PR \cite{lee2007prediction}, PC-SAFT-based formulations \cite{mahmoudabadi2021predictive}, and as molecular descriptors in quantitative structure–property models\cite{niederquell2018new}.

Besides these established applications, $\sigma$-profiles are promising descriptors for machine-learning (ML) models, which have become powerful alternatives to physical models for predicting thermophysical properties \cite{abranches2022sigma,abranches2024stochastic, jirasek2021perspective, jirasek2023combining, hayer2025advancing, hayer2025modified}. The choice of molecular representation is a key factor for ML model performance. A wide range of descriptors has been explored, including coordinate-based atomistic descriptors \cite{rupp2012fast,hansen2015machine,behler2011atom,bartok2013representing}, molecular graphs \cite{schutt2017schnet, duvenaud2015convolutional, hoffmann2025grappa}, text-string encodings (SMILES, SELFIES) \cite{jaeger2018mol2vec,winter2022smile, specht2024hanna, hoffmann2025machine}, and molecular fingerprints \cite{tayyebi2023prediction,xie2020improvement}. The $\sigma$-profile offers a particularly valuable addition to this set, as it provides compact, physics-based information on a molecule’s electronic structure that has proven useful for predicting macroscopic thermodynamic behavior.

Initial studies have shown that incorporating $\sigma$-profiles as molecular descriptors improves the predictive performance of ML models for mixture properties \cite{hayer2025prediction,hayer2025similarity}. However, progress is constrained by the scarcity of publicly available $\sigma$-profiles, as current repositories contain only a few thousand compounds \cite{bell2020benchmark,mullins2006sigma,ferrarini2018open}. Moreover, combining data from different sources is problematic because the underlying quantum-chemical methods used to compute $\sigma$-profiles introduce systematic inconsistencies, limiting comparability. Benchmark studies have demonstrated that changing the quantum-chemistry software, density functional, or basis set systematically shifts the $\sigma$-profile charge density histograms \cite{mu2007performance}, causing energy offsets that persist even after model reparameterization \cite{reinisch2019benchmarking}. Consequently, large-scale $\sigma$-profile collections generated under a single, well-defined quantum-chemical protocol are essential, yet such comprehensive, single-workflow datasets have been lacking.

In this work, we introduce CHAOS (Computed High-Accuracy Observables and Sigma-profiles), a database providing consistent, high-quality $\sigma$-profiles for 53\,091 technically relevant molecules with molecular masses up to 400\,amu, for which experimental thermophysical data are available in the Dortmund Data Bank (DDB) \cite{DortmundDataBank2024}. Beyond $\sigma$-profiles, the database includes a broad range of additional quantum-chemically derived molecular descriptors obtained in a uniform workflow. Table \ref{tab:descriptors} gives an overview of these data types available in CHAOS. Because the database contains complete vibrational and rotational information, derived thermodynamic ideal-gas quantities at various temperatures can also be obtained (see Supporting Information, Section "Calculation of ideal-gas properties").

CHAOS is currently the largest internally consistent database of $\sigma$-profiles and related molecular descriptors, offering a robust foundation for physics-based, data-driven, and hybrid modeling approaches alike.

\begin{table*}[htbp]
\centering
\caption{Overview of quantum-chemical descriptors provided in the CHAOS database. Each descriptor is consistently derived from $\omega$B97X-D/def2-TZVP calculations. For a more detailed overview of all descriptors in the database, please refer to the Supporting Information, Section "Full list of properties within CHAOS".}
\label{tab:descriptors}
\renewcommand{\arraystretch}{1.15}
\begin{tabularx}{\textwidth}{p{3cm} X}
\toprule
\textbf{Category} & \textbf{Descriptors} \\
\midrule

\textbf{Structural} & Optimized Cartesian coordinates \\
 & Rotational constants\\
 & External symmetry number \\
\addlinespace[5pt]

\textbf{Electronic} & Dipole moment vector and magnitude \\
 & Quadrupole tensor and scalar moment \\
 & Polarizability tensor (isotropic and anisotropic) \\
 & Partial charges (Mulliken, APT) \\
 & Energetic gap of frontier orbitals (HOMO--LUMO gap) \\
 & Self-consistent field energy (SCF energy) \\

\textbf{Vibrational} & Harmonic frequencies, IR intensities, reduced masses, and force constants \\
 & Zero-point energy (ZPE) \\
 & Ideal-gas isochoric heat capacitiy and entropy \\
\addlinespace[5pt]

\textbf{NMR} & Isotropic and anisotropic shielding constants \\
 & Dia-/paramagnetic susceptibility tensors \\
\addlinespace[5pt]

\textbf{Solvation} & Cavity information \\
 & Per-atom surface area and charge \\
 & Electrostatic and non-electrostatic solvation energies \\
 & $\sigma$-profiles (total and partial) \\

\bottomrule
\end{tabularx}
\end{table*}

\section{Workflow and methods}
\label{sec:workflow}
\begin{figure}[H]
    \centering
    \includegraphics[width=\textwidth]{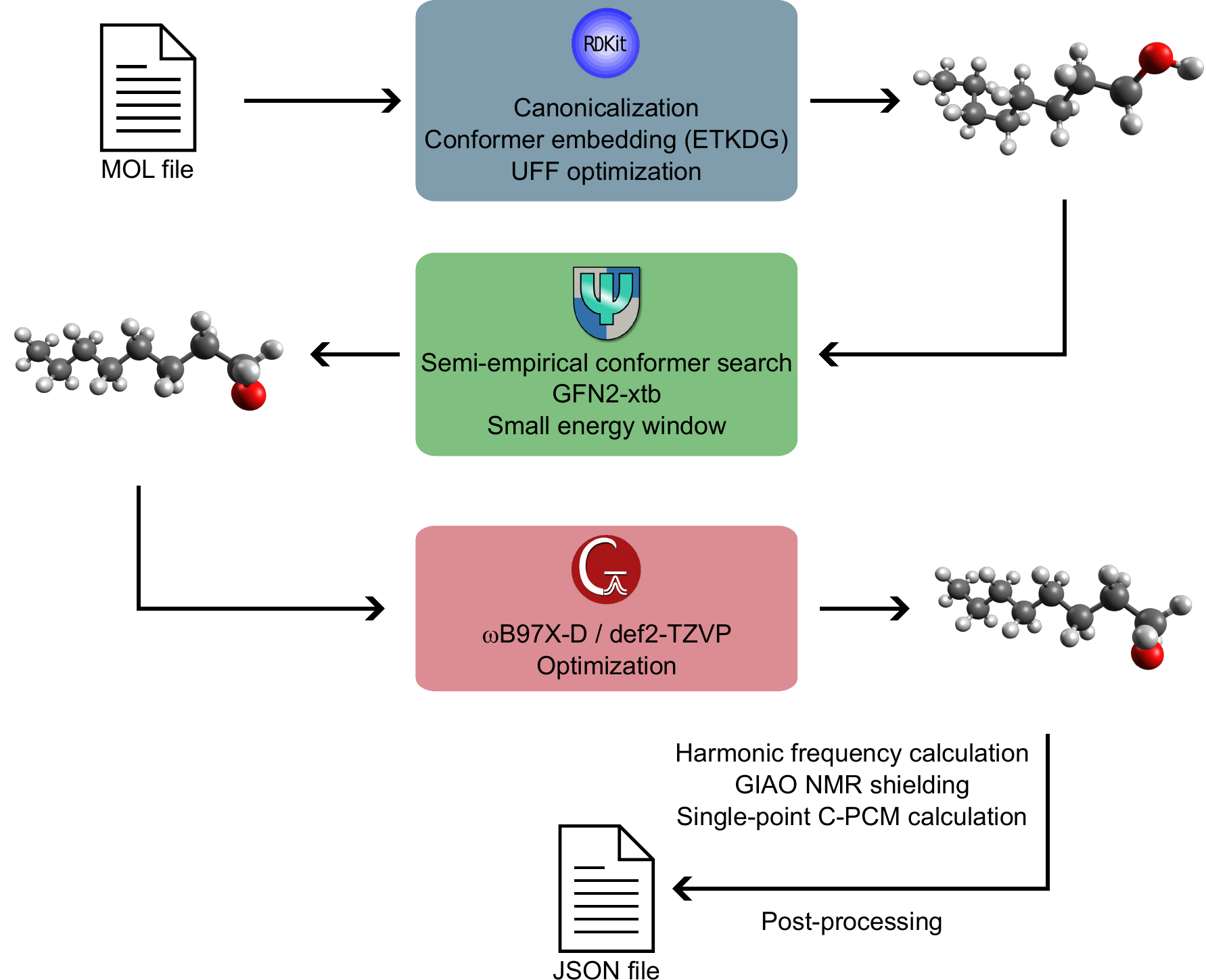}
    \caption{Schematic overview of the CHAOS data generation workflow. Starting from MOL files, defining a non-optimized geometry, molecules undergo canonicalization and 3D conformer embedding via RDKit \cite{rdkit} using the ETKDG algorithm \cite{riniker2015better}, followed by a geometry optimization based on a universal force field (UFF)\cite{rappe1992uff} (top). The lowest-energy structure is subsequently refined via a semi-empirical conformer search using CREST \cite{pracht2020automated} with GFN2-xTB \cite{bannwarth2019gfn2} in a narrow energy window (middle). The final high-level geometry optimization is performed using Gaussian 16 \cite{g16} at the $\omega$B97X-D/def2-TZVP \cite{chai2008systematic,weigend2005balanced} level of density functional theory (DFT) \cite{kohn1965self}. Based on this geometry, subsequent calculations were performed, including harmonic frequency analysis, gauge‑including atomic orbital (GIAO) nuclear magnetic resonance  (NMR) shielding computations\cite{ruud1993hartree}, and single-point conductor-like polarizable continuum (C-PCM) calculations \cite{barone1998quantum,cossi2003energies} (bottom). The resulting data are further postprocessed to derive molecular descriptors, such as the $\sigma$-profile. All results were compiled into structured JSON records.}
    \label{fig:workflow}
\end{figure}

Figure \ref{fig:workflow} gives an overview of the workflow used for data generation in this work. The details of the individual steps are described in the following sections.

\subsection{Generation of initial molecular geometries}
\label{sub:rdkit}

The generation of the initial geometries followed a standardized multistep procedure to ensure consistent and physically meaningful 3D structures for all molecules. For each component listed in the DDB \cite{DortmundDataBank2024}, the initial molecular structure was imported from a MOL file and converted to a canonical representation to remove inconsistencies in atom ordering and bonding information. Three-dimensional conformers were then generated by distance-geometry embedding, followed by an initial energy prescreening using a generic molecular-mechanics force field. From the resulting conformer ensemble, redundant or high-energy structures were discarded, leaving only low-energy, thermally accessible candidates. The lowest-energy conformer from this filtered set was reoptimized to yield the structure used for further optimization.

All operations were carried out with RDKit \cite{rdkit}, employing the ETKDG\cite{riniker2015better} embedding algorithm and the universal force field (UFF)\cite{rappe1992uff}. Up to 300 conformers were generated per molecule, and those differing by less than 1\,kcal\,mol$^{-1}$ in energy were removed to eliminate near-degenerate geometries. The procedure efficiently produced representative, low-energy gas-phase structures while avoiding unnecessary computational overhead in later semi-empirical optimizations. Implementation-level details, including function calls and script examples, are provided in the Supporting Information.

\subsection{Semi‐empirical conformer refinement}
\label{sub:crest}
The lowest-energy conformer identified in the force-field search served as the starting point for an exhaustive conformational search. This step bridges the gap between classical molecular mechanics and full density functional theory (DFT) \cite{kohn1965self}, ensuring that all molecules enter the high-level calculations starting from physically meaningful geometries.

The search was carried out using CREST (conformer–rotamer sampling tool)\cite{pracht2020automated} using the GFN2-xTB\cite{bannwarth2019gfn2} (geometry, frequency, non-covalent, extended tight binding) Hamiltonian. This method includes an explicit electronic treatment of valence electrons and thereby captures non-covalent interactions -- most importantly hydrogen bonding, dispersion, and hyperconjugation -- that are not accounted for by generic force-field methods. This procedure systematically improves molecular geometries while maintaining computational feasibility for large datasets.

To balance accuracy and cost, the search was restricted to conformers within a narrow energy window, retaining only thermally relevant minima. From this ensemble, the conformer of lowest energy was selected and forwarded to the subsequent DFT calculations described below. Further computational parameters and an example input are provided in the Supporting Information.

\subsection{High‐level quantum‐chemical calculations (Gaussian\,16)}
\label{sub:gaussian}

Starting from the semi-empirical geometries obtained by the steps described above, all molecules were treated with a single, uniform DFT\cite{kohn1965self} workflow in Gaussian 16\cite{g16}. The $\omega$B97X-D functional\cite{chai2008systematic} in combination with the def2-TZVP (triple-$\zeta$) basis set\cite{weigend2005balanced} was used throughout to ensure that structures and properties are directly comparable across all 53\,091 components. After optimization, a harmonic frequency analysis was performed to confirm that the stationary points are true minima with no imaginary modes. From the same calculation, we obtained the electronic and vibrational observables used in CHAOS, ensuring that spectra and electronic properties are tied to the converged electronic structures. Entries with a remaining imaginary mode after a reoptimization are transparently flagged in the data release. Subsequently, gauge‑including atomic orbital (GIAO) nuclear magnetic resonance (NMR) shielding tensors \cite{ruud1993hartree} were computed, from which the NMR observables (per-atom isotropic shielding and anisotropy, magnetic susceptibilities) were obtained. Finally, a single-point conductor-like polarizable continuum (C-PCM)\cite{barone1998quantum,cossi2003energies} calculation was performed on each optimized gas-phase minimum geometry. This yields the tessellated COSMO surface and per-segment data (positions, areas, surface charge densities, potentials) together with the associated energetic terms. These raw segments are released so users can construct $\sigma$-profiles with any postprocessing they prefer; for convenience, we also provide precomputed $\sigma$-profiles following the open-source COSMO-SAC-dsp protocol\cite{bell2020benchmark}. Exact keywords and example Gaussian 16 inputs are documented in the Supporting Information. Open-source alternatives for these steps exist; for example, COSMO calculations and the generation of $\sigma$-profiles can be carried out with NWChem \cite{apra2020nwchem, soares2025recent} and openCOSMO-RS \cite{gerlach2022open}. We therefore cite these approaches as an accessible alternative for community-driven extensions.

\section{Data records}
\label{sec:datarecords}

CHAOS provides a self-contained dataset for each molecule, structured into six primary categories: \texttt{general}, \texttt{structural}, \texttt{electronic}, \texttt{vibrational}, \texttt{NMR}, and \texttt{solvation}. Each molecular entry is stored in a single JSON file accompanied by an internal identifier and canonical SMILES string. The complete list of field names, data types, and units is given in the Supporting Information (“Full list of properties within CHAOS”). In addition, the Supporting Information includes a validation on a small representative subset of molecules, comparing unscaled zero-point energies against the NIST CCCBDB \cite{johnson1998nist} reference database, together with summary error metrics.

The \textbf{general} block contains metadata and identifiers: the canonical SMILES string, the molecular formula and molecular mass, and a complete atom register that maps every internal atom index to its element symbol and atomic number. This register serves as a reference backbone for all per-atom arrays in subsequent sections. A Boolean flag (\texttt{not\_converged}) marks molecules for which an imaginary frequency persisted after reoptimization, allowing users to filter or handle such cases explicitly.

The \textbf{structural} block stores Cartesian coordinates of all atoms in the molecule as an \(N\times3\) matrix in units of \AA, ordered consistently with the atom register. It also includes rotational constants $A$, $B$, and $C$ (or a single constant for linear species) and the external symmetry number, all derived from the harmonic frequency analysis.

The \textbf{electronic} block collects quantities obtained directly from the converged Kohn–Sham wavefunction. It includes the vectorial and scalar dipole moments, the full quadrupole tensor, its moment \cite{gray1984tmf} and its traceless form, dipole polarizabilities (tensor, isotropic, and anisotropic), and per-atom partial charges computed by both the Mulliken\cite{mulliken1955electronic} and atomic polar tensor (APT)\cite{person1974dipole} scheme.  For heavy-atom analyses, hydrogen charges are summed onto the bonded heavy atoms, providing an optional condensed charge representation. Additionally, the final self-consistent field (SCF) energy and the HOMO–LUMO gap (energy difference of frontier orbitals) are reported.

The \textbf{vibrational} block provides a complete record of the harmonic frequency calculation: all wavenumbers $\tilde{\nu}_i$, IR intensities, force constants, and reduced masses, each stored as lists of length $3N-F$, where $F$ is the number of rotational degrees of freedom ($F=5$ for linear and $F=6$ for non-linear molecules). The zero-point energy is calculated according to Equation (\ref{eq:ZPE})

\begin{equation}
\label{eq:ZPE}
    E_\mathrm{ZPE}= \frac{1}{2}\sum_i^{3N-F} hc\tilde{\nu}_i
\end{equation}

and reported in Hartree. For structures flagged \texttt{non\_converged}, negative frequencies are retained with a negative sign to preserve internal consistency. Moreover, the ideal-gas isochoric heat capacities and the molecules' entropy at \SI{0}{\kelvin} are provided.

The \textbf{NMR} block includes isotropic and anisotropic magnetic shieldings for every nucleus, as well as dia- and paramagnetic susceptibility tensors. These data permit the computation of derived magnetic properties, such as bulk susceptibility or chemical shielding in solution.

The \textbf{solvation} block assembles all output from the single-point C-PCM calculation. It reports the cavity surface area $A_\sigma$, cavity volume $V_\sigma$, and number of tessellated surface segments $N_\mathrm{seg}$. For each atom, the associated surface area, average surface charge density, and COSMO-derived atomic charge are stored. Energetic components include the self-consistent COSMO energy, the dielectric correction, and the non-electrostatic contributions -- cavitation, dispersion, and Pauli repulsion -- whose sum yields the total non-electrostatic energy. Finally, the complete set of surface segments is provided as a nested list, where each entry contains the segment index, parent atom index, Cartesian coordinates, segment charge, area, surface charge density, and potential. These data form the raw basis for constructing $\sigma$-profiles. For convenience, CHAOS also provides $\sigma$-profiles pre-computed according to the open-source COSMO-SAC-dsp protocol \cite{bell2020benchmark}, subdivided into three partial contributions: non-hydrogen-bonding (NHB), hydrogen bonds of a hydroxyl group (OH), and other hydrogen bonds, i.e., oxygen (e.g., within ketones, aldehydes, ethers), nitrogen, fluorine, and hydrogen bonded to nitrogen or fluorine (OT). Each profile is tabulated over segment charge densities from -0.025 to 0.025 $e$/\AA\:in steps of 0.001 $e$/\AA.

\section{Overview of CHAOS database}
\begin{figure}[H]
    \centering
    \includegraphics[width=\textwidth]{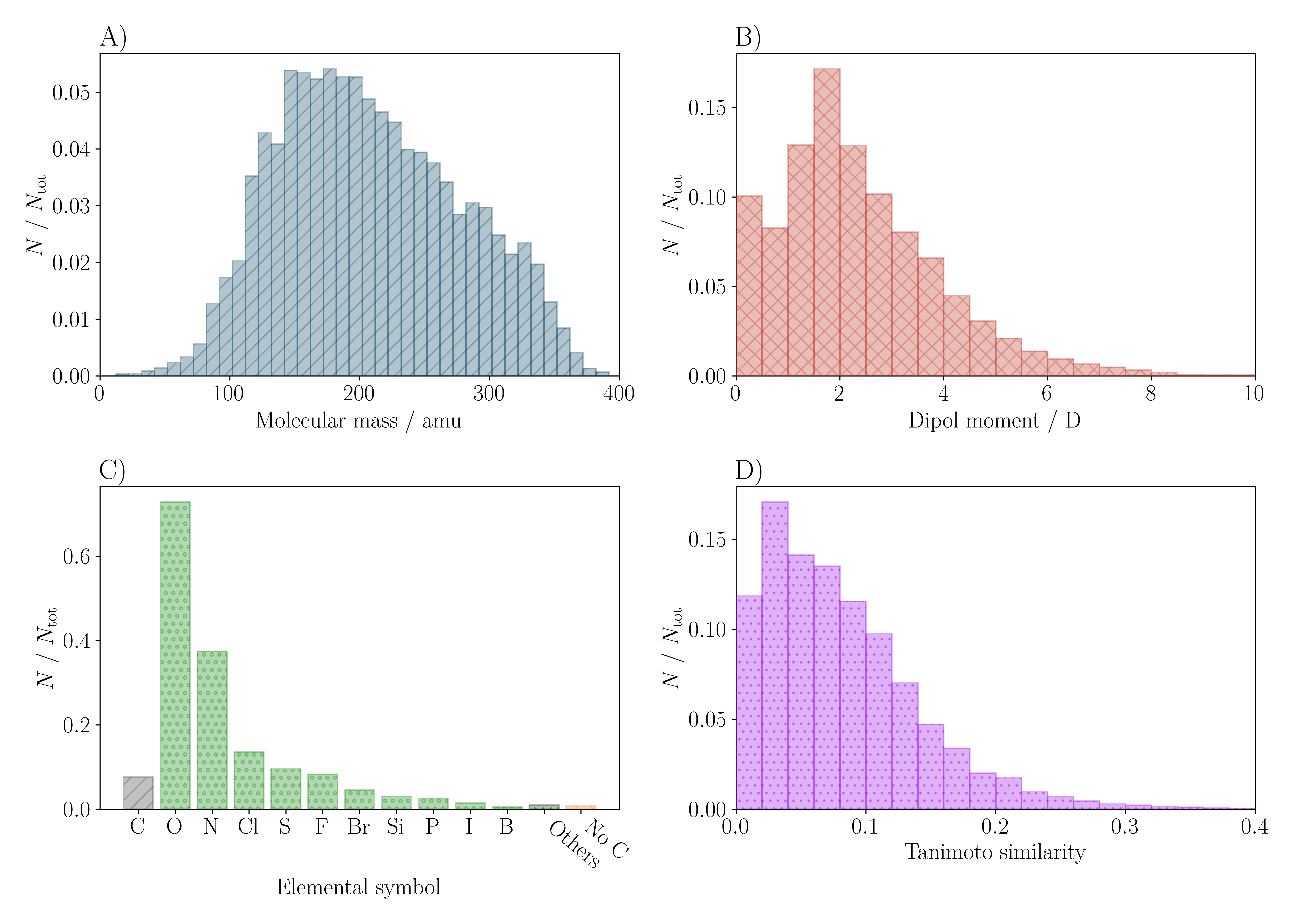}
    \caption{Overview of the diversity of molecules included in the CHAOS database. Panels A) and B) show the distributions of the molecular masses and dipole moments, respectively.  Panel C) gives the relative numbers of molecules containing specific heteroatoms (light green bars) together with the relative number of molecules containing other heteroatoms (dark green bar). Additionally, the number of hydrocarbons is reported (grey bar), as well as the number of molecules containing no carbon (orange). The number of molecules in a given class is reported as a fraction of the total number of molecules $N_\mathrm{tot} =$ 53\,091 in Panels A) - C). Panel D) shows the distribution of pairwise Tanimoto similarities\cite{bajusz2015tanimoto} (ECFP4\cite{rogers2010extended,morgan1965generation}, 2048 bits) calculated for a random sample of $10^7$ molecule pairs from CHAOS; \SI{99.80}{\percent} of the reported similarities are within the shown range.}
    \label{fig:tech_val}
\end{figure}

In Figure \ref{fig:tech_val}, some characteristics of the molecules included in the CHAOS database are shown, namely, distributions with regard to molecular mass, molecular polarity, elemental composition, and structural similarity between pairs of components calculated via the Tanimoto similarity. Further statistics and additional figures are provided in the Supporting Information.

The molecular masses range from 2\,amu for hydrogen to 394\,amu for iodoform, cf. Figure \ref{fig:tech_val} A). The dipole moments (calculated for the gas phase) range from 0\,D for hydrocarbons to 15.1 D for $\alpha$-2'-deoxythioguanosine. In calculating the number of molecules with heteroatoms for the plot shown in Figure \ref{fig:tech_val} C), molecules were counted if they contain at least one heteroatom of the respective type. The vast majority of the molecules in CHAOS are organic, as can be seen from the small bar for molecules containing no carbon atoms in Figure \ref{fig:tech_val} C). About \SI{7}{\percent} of the molecules are hydrocarbons. Oxygen dominates the heteroatom landscape, followed by nitrogen, sulfur, and halogens. Finally, internal diversity was quantified using the Tanimoto similarity\cite{bajusz2015tanimoto}, which measures the structural resemblance between two molecules as the ratio of shared to total features encoded in their molecular fingerprints, such as atom environments or substructural patterns. The Tanimoto similarity ranges from 0 (no similarity) to 1 (identical). A value of 0.6 is often taken as the threshold for considering two molecules to be similar \cite{huber2021ms2deepscore, blaschke2020memory}. A total of $10^7$ pairs of randomly selected molecules from CHAOS were considered for the similarity analysis shown in Figure \ref{fig:tech_val} D). The results demonstrate that the vast majority of pairs of molecules in CHAOS show little similarity, underpinning the diversity of the CHAOS dataset.

\section{Comparison with existing databases}
\label{sec:comparison}

Until now, the three largest openly accessible collections of $\sigma$-profiles have been the LVPP set\cite{ferrarini2018open}, the Virginia Tech VT-2005 library\cite{mullins2006sigma}, and the NIST COSMO-SAC files\cite{bell2020benchmark}, providing data for 1\,755, 1\,432, and 2\,261 molecules, respectively. With 53\,091 components, CHAOS substantially expands the publicly available chemical space for $\sigma$-profiles. Moreover, while a combination of the previously available databases would entail mixed quantum-chemical levels -- Hartree–Fock (HF)/TZVP (triple-$\zeta$) for LVPP and GGA-VWN-BP/DNP (double-$\zeta$) for the others -- CHAOS applies a single $\omega$B97X-D/def2-TZVP protocol throughout, rendering all solvation descriptors directly comparable. In addition, the release of raw C-PCM output files allows users to regenerate $\sigma$-profiles using their own postprocessing procedures.

A similar pattern emerges for vibrational information. The NIST CCCBDB archive (2\,186 molecules, computed with methods ranging from low to high level)\cite{johnson1998nist} provides partially benchmark-quality frequencies but spans only a limited chemical space and includes no solvation descriptors. The recently released Hessian QM9 extension provides full Hessians and normal modes for 41\,645 molecules at the $\omega$B97X/6-31G* (double-$\zeta$) level in four implicit solvents,\cite{williams2025hessian}, yet still omits explicit COSMO data. CHAOS exceeds Hessian QM9 in both methodological rigor and property scope; however, its vibrational frequencies are available only for the gas phase, not for condensed phases, making the two databases complementary rather than directly competitive.

Larger quantum-chemical corpora, such as QM7-X\cite{hoja2021qm7} (42 million molecules at approximately triple-$\zeta$ level of theory) and QM9 \cite{ramakrishnan2014quantum} (133\,885 molecules at B3LYP/6-31G* double-$\zeta$ quality), primarily target energies, forces, and electronic properties while omitting $\sigma$-profiles and IR/NMR observables. CHAOS therefore occupies an intermediate niche: although its 53\,091 molecules cover a smaller chemical space than QM9, each structure is treated consistently at a triple-$\zeta$ level ($\omega$B97X-D/def2-TZVP), thus surpassing QM9 in methodological accuracy. Beyond energies and standard electronic descriptors, CHAOS uniquely integrates gas-phase vibrational spectra, GIAO NMR tensors, and COSMO-based solvation surfaces with $\sigma$-profiles. This unified set of NMR, IR, solvation, and electronic data positions CHAOS as a complementary resource rather than a direct competitor to the available very large, energy-centric databases.

\section{Conclusions}
\label{sec:conclusion}

We have introduced CHAOS, a comprehensive quantum-chemical database that provides complete outputs from C-PCM calculations, enabling the generation of custom $\sigma$-profiles together with precomputed COSMO-SAC-dsp $\sigma$-profiles, IR spectra, ideal-gas heat capacities and entropies, GIAO NMR tensors, and a broad suite of electronic descriptors for 53\,091 molecules. All data were obtained using a uniform $\omega$B97X-D/def2-TZVP protocol encompassing geometry optimizations, harmonic frequency analyses, GIAO calculations, and C-PCM single-point evaluations. This consistent methodology ensures full comparability across the entire dataset. The broad structural diversity of CHAOS, spanning a wide range of molecular sizes, polarities, and functional groups, makes it particularly suitable for benchmarking and for training data-driven thermodynamic models.

In comparison with existing resources, CHAOS expands the publicly available $\sigma$-profile space by more than an order of magnitude relative to the previously largest collections -- LVPP\cite{ferrarini2018open}, VT-2005\cite{mullins2006sigma}, and the NIST COSMO-SAC files\cite{bell2020benchmark} -- while avoiding the protocol heterogeneity that otherwise complicates the consistent aggregation of $\sigma$-profiles across different sources. It further links harmonic vibrational data with solvation descriptors, a combination not offered by either the high-accuracy but limited CCCBDB\cite{johnson1998nist} or the broader Hessian QM9 dataset.\cite{williams2025hessian} Within the wider landscape of quantum-chemical corpora, CHAOS bridges the gap between large-scale, energy-focused databases, such as QM9\cite{ramakrishnan2014quantum} and QM7-X\cite{hoja2021qm7}, and the property-oriented needs of chemical engineering and solvent design.

By uniting gas-phase quantum-chemical data with consistent, ready-to-use $\sigma$-profiles and spectroscopic observables, CHAOS provides a coherent foundation for developing, training, and benchmarking machine-learning models in thermodynamic property prediction. The database also facilitates the reparameterization and validation of COSMO-based models and offers a versatile resource for algorithmic screening, solvent design, and multiscale process modeling. Future extensions will aim to expand CHAOS to include condensed-phase and reactive systems, integrate additional solvation models, and enhance data accessibility for community-driven reuse.

The CHAOS database is freely available under an open license, providing an extensible platform for advancing data-driven modeling of physicochemical properties.

\section{Data and software availability}
\label{sec:dataavail}
The full CHAOS database is openly available at Zenodo:

\begin{center}
\url{https://doi.org/10.5281/zenodo.17691924}
\end{center}

The repository includes a ZIP file containing 53\,091 JSON files, one for each respective molecule, as well as one JSON dictionary assigning the database ID and the SMILES of the respective molecule for a better accessibility. Moreover, an associated GitHub repository with the used workflow is available at:
\begin{center}
\url{https://github.com/GondLTD/CHAOS_database}
\end{center}

All data may be used, modified, and redistributed under the terms of the CC-BY-4.0 licence.

\section*{Acknowledgment}
We gratefully acknowledge financial support by the Carl Zeiss Foundation in the frame of the project 'Process Engineering 4.0' and by DFG in the frame of the Priority Program SPP2363 'Molecular Machine Learning' (grant number 497201843). Furthermore, FJ gratefully acknowledges financial support by DFG in the frame of the Emmy-Noether program (grant number 528649696). The simulations were carried out on the HPC machine ELWE at the RHRZ under the grant RPTU-MLVT.

\section*{Supporting Information}
Full list of CHAOS database fields with descriptions, data types, and units; Distributions of scalar properties (histograms) for the complete dataset; Detailed derivation of ideal-gas properties within the rigid-rotor harmonic-oscillator (RRHO) approximation; Implementation details of the workflow, including RDKit-based canonicalization, 3D conformer generation and filtering, CREST conformer refinement, and Gaussian 16 input templates for optimization, frequency, GIAO NMR, and C-PCM calculations; Validation of unscaled zero-point energies and dipole moments against CCCBDB reference data.

\clearpage
\bibliography{literature.bib}

@article{klamt1995conductor,
  title={Conductor-like screening model for real solvents: a new approach to the quantitative calculation of solvation phenomena},
  author={Klamt, Andreas},
  journal={The Journal of Physical Chemistry},
  volume={99},
  number={7},
  pages={2224--2235},
  year={1995},
  publisher={ACS Publications}
}

@article{winter2022smile,
  title={A smile is all you need: predicting limiting activity coefficients from SMILES with natural language processing},
  author={Winter, Benedikt and Winter, Clemens and Schilling, Johannes and Bardow, Andr{\'e}},
  journal={Digital Discovery},
  volume={1},
  number={6},
  pages={859--869},
  year={2022},
  publisher={Royal Society of Chemistry}
}

@article{specht2024hanna,
  title={HANNA: hard-constraint neural network for consistent activity coefficient prediction},
  author={Specht, Thomas and Nagda, Mayank and Fellenz, Sophie and Mandt, Stephan and Hasse, Hans and Jirasek, Fabian},
  journal={Chemical Science},
  volume={15},
  number={47},
  pages={19777--19786},
  year={2024},
  publisher={Royal Society of Chemistry}
}

@article{hoffmann2025grappa,
  title={GRAPPA--A Hybrid Graph Neural Network for Predicting Pure Component Vapor Pressures},
  author={Hoffmann, Marco and Hasse, Hans and Jirasek, Fabian},
  journal={arXiv preprint arXiv:2501.08729},
  year={2025}
}

@article{rupp2012fast,
  title={Fast and accurate modeling of molecular atomization energies with machine learning},
  author={Rupp, Matthias and Tkatchenko, Alexandre and M{\"u}ller, Klaus-Robert and Von Lilienfeld, O Anatole},
  journal={Physical review letters},
  volume={108},
  number={5},
  pages={058301},
  year={2012},
  publisher={APS}
}

@article{hansen2015machine,
  title={Machine learning predictions of molecular properties: Accurate many-body potentials and nonlocality in chemical space},
  author={Hansen, Katja and Biegler, Franziska and Ramakrishnan, Raghunathan and Pronobis, Wiktor and Von Lilienfeld, O Anatole and Muller, Klaus-Robert and Tkatchenko, Alexandre},
  journal={The journal of physical chemistry letters},
  volume={6},
  number={12},
  pages={2326--2331},
  year={2015},
  publisher={ACS Publications}
}

@article{behler2011atom,
  title={Atom-centered symmetry functions for constructing high-dimensional neural network potentials},
  author={Behler, J{\"o}rg},
  journal={The Journal of chemical physics},
  volume={134},
  number={7},
  year={2011},
  publisher={AIP Publishing}
}

@article{bartok2013representing,
  title={On representing chemical environments},
  author={Bart{\'o}k, Albert P and Kondor, Risi and Cs{\'a}nyi, G{\'a}bor},
  journal={Physical Review B—Condensed Matter and Materials Physics},
  volume={87},
  number={18},
  pages={184115},
  year={2013},
  publisher={APS}
}

@article{rogers2010extended,
  title={Extended-connectivity fingerprints},
  author={Rogers, David and Hahn, Mathew},
  journal={Journal of chemical information and modeling},
  volume={50},
  number={5},
  pages={742--754},
  year={2010},
  publisher={ACS Publications}
}

@article{abranches2022sigma,
  title={Sigma profiles in deep learning: towards a universal molecular descriptor},
  author={Abranches, Dinis O and Zhang, Yong and Maginn, Edward J and Col{\'o}n, Yamil J},
  journal={Chemical Communications},
  volume={58},
  number={37},
  pages={5630--5633},
  year={2022},
  publisher={Royal Society of Chemistry}
}

@article{jaeger2018mol2vec,
  title={Mol2vec: unsupervised machine learning approach with chemical intuition},
  author={Jaeger, Sabrina and Fulle, Simone and Turk, Samo},
  journal={Journal of chemical information and modeling},
  volume={58},
  number={1},
  pages={27--35},
  year={2018},
  publisher={ACS Publications}
}

@book{klamt2005cosmo,
  title={COSMO-RS: from quantum chemistry to fluid phase thermodynamics and drug design},
  author={Klamt, Andreas},
  year={2005},
  publisher={Elsevier}
}

@article{bell2020benchmark,
  title={A benchmark open-source implementation of COSMO-SAC},
  author={Bell, Ian H and Mickoleit, Erik and Hsieh, Chieh-Ming and Lin, Shiang-Tai and Vrabec, Jadran and Breitkopf, Cornelia and J{\"a}ger, Andreas},
  journal={Journal of chemical theory and computation},
  volume={16},
  number={4},
  pages={2635--2646},
  year={2020},
  publisher={ACS Publications}
}

@article{mullins2006sigma,
  title={Sigma-profile database for using COSMO-based thermodynamic methods},
  author={Mullins, Eric and Oldland, Richard and Liu, YA and Wang, Shu and Sandler, Stanley I and Chen, Chau-Chyun and Zwolak, Michael and Seavey, Kevin C},
  journal={Industrial \& engineering chemistry research},
  volume={45},
  number={12},
  pages={4389--4415},
  year={2006},
  publisher={ACS Publications}
}

@article{ferrarini2018open,
  title={An open and extensible sigma-profile database for COSMO-based models},
  author={Ferrarini, F and Fl{\^o}res, GB and Muniz, AR and de Soares, RP},
  journal={AIChE Journal},
  volume={64},
  number={9},
  pages={3443--3455},
  year={2018},
  publisher={Wiley Online Library}
}

@misc{DortmundDataBank2024,
  author = {{Dortmund Data Bank}},
  title = {{, 2024}},
  howpublished = {\url{www.ddbst.com}}
}

@misc{rdkit,
  author       = {Landrum, Gregory},
  title        = {{RDKit}: Open-source cheminformatics},
  howpublished = {\url{https://www.rdkit.org}},
  note         = {Accessed: 2025-07-10}
}

@article{riniker2015better,
  title={Better informed distance geometry: using what we know to improve conformation generation},
  author={Riniker, Sereina and Landrum, Gregory A},
  journal={Journal of chemical information and modeling},
  volume={55},
  number={12},
  pages={2562--2574},
  year={2015},
  publisher={ACS Publications}
}

@article{rappe1992uff,
  title={UFF, a full periodic table force field for molecular mechanics and molecular dynamics simulations},
  author={Rapp{\'e}, Anthony K and Casewit, Carla J and Colwell, KS and Goddard III, William A and Skiff, W Mason},
  journal={Journal of the American chemical society},
  volume={114},
  number={25},
  pages={10024--10035},
  year={1992},
  publisher={ACS Publications}
}

@article{pracht2020automated,
  title={Automated exploration of the low-energy chemical space with fast quantum chemical methods},
  author={Pracht, Philipp and Bohle, Fabian and Grimme, Stefan},
  journal={Physical Chemistry Chemical Physics},
  volume={22},
  number={14},
  pages={7169--7192},
  year={2020},
  publisher={Royal Society of Chemistry}
}

@article{bannwarth2019gfn2,
  title={GFN2-xTB—An accurate and broadly parametrized self-consistent tight-binding quantum chemical method with multipole electrostatics and density-dependent dispersion contributions},
  author={Bannwarth, Christoph and Ehlert, Sebastian and Grimme, Stefan},
  journal={Journal of chemical theory and computation},
  volume={15},
  number={3},
  pages={1652--1671},
  year={2019},
  publisher={ACS Publications}
}

@article{kohn1965self,
  title={Self-consistent equations including exchange and correlation effects},
  author={Kohn, Walter and Sham, Lu Jeu},
  journal={Physical review},
  volume={140},
  number={4A},
  pages={A1133},
  year={1965},
  publisher={APS}
}

@article{chai2008systematic,
  title={Systematic optimization of long-range corrected hybrid density functionals},
  author={Chai, Jeng-Da and Head-Gordon, Martin},
  journal={The Journal of chemical physics},
  volume={128},
  number={8},
  year={2008},
  publisher={AIP Publishing}
}

@article{weigend2005balanced,
  title={Balanced basis sets of split valence, triple zeta valence and quadruple zeta valence quality for H to Rn: Design and assessment of accuracy},
  author={Weigend, Florian and Ahlrichs, Reinhart},
  journal={Physical Chemistry Chemical Physics},
  volume={7},
  number={18},
  pages={3297--3305},
  year={2005},
  publisher={Royal Society of Chemistry}
}

@misc{g16,
author={M. J. Frisch and G. W. Trucks and H. B. Schlegel and G. E. Scuseria and M. A. Robb and J. R. Cheeseman and G. Scalmani and V. Barone and G. A. Petersson and H. Nakatsuji and X. Li and M. Caricato and A. V. Marenich and J. Bloino and B. G. Janesko and R. Gomperts and B. Mennucci and H. P. Hratchian and J. V. Ortiz and A. F. Izmaylov and J. L. Sonnenberg and D. Williams-Young and F. Ding and F. Lipparini and F. Egidi and J. Goings and B. Peng and A. Petrone and T. Henderson and D. Ranasinghe and V. G. Zakrzewski and J. Gao and N. Rega and G. Zheng and W. Liang and M. Hada and M. Ehara and K. Toyota and R. Fukuda and J. Hasegawa and M. Ishida and T. Nakajima and Y. Honda and O. Kitao and H. Nakai and T. Vreven and K. Throssell and Montgomery, {Jr.}, J. A. and J. E. Peralta and F. Ogliaro and M. J. Bearpark and J. J. Heyd and E. N. Brothers and K. N. Kudin and V. N. Staroverov and T. A. Keith and R. Kobayashi and J. Normand and K. Raghavachari and A. P. Rendell and J. C. Burant and S. S. Iyengar and J. Tomasi and M. Cossi and J. M. Millam and M. Klene and C. Adamo and R. Cammi and J. W. Ochterski and R. L. Martin and K. Morokuma and O. Farkas and J. B. Foresman and D. J. Fox},
title={Gaussian˜16 {R}evision {C}.01},
year={2016},
note={Gaussian Inc. Wallingford CT}
}

@article{ruud1993hartree,
  title={Hartree--Fock limit magnetizabilities from London orbitals},
  author={Ruud, Kenneth and Helgaker, Trygve and Bak, Keld L and Jo/rgensen, Poul and Jensen, Hans Jo/rgen Aa},
  journal={The Journal of chemical physics},
  volume={99},
  number={5},
  pages={3847--3859},
  year={1993},
  publisher={American Institute of Physics}
}

@article{barone1998quantum,
  title={Quantum calculation of molecular energies and energy gradients in solution by a conductor solvent model},
  author={Barone, Vincenzo and Cossi, Maurizio},
  journal={The Journal of Physical Chemistry A},
  volume={102},
  number={11},
  pages={1995--2001},
  year={1998},
  publisher={ACS Publications}
}

@article{cossi2003energies,
  title={Energies, structures, and electronic properties of molecules in solution with the C-PCM solvation model},
  author={Cossi, Maurizio and Rega, Nadia and Scalmani, Giovanni and Barone, Vincenzo},
  journal={Journal of computational chemistry},
  volume={24},
  number={6},
  pages={669--681},
  year={2003},
  publisher={Wiley Online Library}
}

@article{bajusz2015tanimoto,
  title={Why is Tanimoto index an appropriate choice for fingerprint-based similarity calculations?},
  author={Bajusz, D{\'a}vid and R{\'a}cz, Anita and H{\'e}berger, K{\'a}roly},
  journal={Journal of cheminformatics},
  volume={7},
  pages={1--13},
  year={2015},
  publisher={Springer}
}

@article{morgan1965generation,
  title={The generation of a unique machine description for chemical structures-a technique developed at chemical abstracts service.},
  author={Morgan, Harry L},
  journal={Journal of chemical documentation},
  volume={5},
  number={2},
  pages={107--113},
  year={1965},
  publisher={ACS Publications}
}

@misc{johnson1998nist,
  title={NIST computational chemistry comparison and benchmark database},
  author={Johnson, Russell D and others},
  year={1998},
  publisher={American Institute of Chemical Engineers}
}

@article{ramakrishnan2014quantum,
  title={Quantum chemistry structures and properties of 134 kilo molecules},
  author={Ramakrishnan, Raghunathan and Dral, Pavlo O and Rupp, Matthias and Von Lilienfeld, O Anatole},
  journal={Scientific data},
  volume={1},
  number={1},
  pages={1--7},
  year={2014},
  publisher={Nature Publishing Group}
}

@article{hoja2021qm7,
  title={QM7-X, a comprehensive dataset of quantum-mechanical properties spanning the chemical space of small organic molecules},
  author={Hoja, Johannes and Medrano Sandonas, Leonardo and Ernst, Brian G and Vazquez-Mayagoitia, Alvaro and DiStasio Jr, Robert A and Tkatchenko, Alexandre},
  journal={Scientific data},
  volume={8},
  number={1},
  pages={43},
  year={2021},
  publisher={Nature Publishing Group UK London}
}

@article{williams2025hessian,
  title={Hessian QM9: A quantum chemistry database of molecular Hessians in implicit solvents},
  author={Williams, Nicholas J and Kabalan, Lara and Stojanovic, Ljiljana and Z{\'o}lyomi, Viktor and Pyzer-Knapp, Edward O},
  journal={Scientific data},
  volume={12},
  number={1},
  pages={9},
  year={2025},
  publisher={Nature Publishing Group UK London}
}

@book{gray1984tmf,
  author       = {Gray, C. G. and Gubbins, K. E.},
  title        = {Theory of Molecular Fluids: {I}. Fundamentals},
  series       = {International Series of Monographs on Chemistry},
  volume       = {1},
  publisher    = {Oxford University Press},
  address      = {Oxford},
  year         = {1984},
  edition      = {Illustrated, new ed.},
  isbn         = {9780198556022},
  pages        = {626},
  note         = {Oxford Science Publications}
}

@article{hayer2025prediction,
  title={Prediction of activity coefficients by similarity-based imputation using quantum-chemical descriptors},
  author={Hayer, Nicolas and Specht, Thomas and Arweiler, Justus and Gond, Dominik and Hasse, Hans and Jirasek, Fabian},
  journal={Physical Chemistry Chemical Physics},
  volume={27},
  number={8},
  pages={4307--4315},
  year={2025},
  publisher={Royal Society of Chemistry}
}

@article{hayer2025similarity,
  title={Similarity-Informed Matrix Completion Method for Predicting Activity Coefficients},
  author={Hayer, Nicolas and Specht, Thomas and Arweiler, Justus and Hasse, Hans and Jirasek, Fabian},
  journal={The Journal of Physical Chemistry A},
  volume={129},
  number={13},
  pages={3141--3147},
  year={2025},
  publisher={ACS Publications}
}

@article{lin2002priori,
  title={A priori phase equilibrium prediction from a segment contribution solvation model},
  author={Lin, Shiang-Tai and Sandler, Stanley I},
  journal={Industrial \& engineering chemistry research},
  volume={41},
  number={5},
  pages={899--913},
  year={2002},
  publisher={ACS Publications}
}

@article{reinisch2019benchmarking,
  title={Benchmarking different QM levels for usage with COSMO-RS},
  author={Reinisch, Jens and Diedenhofen, Michael and Wilcken, Rainer and Udvarhelyi, Anik{\'o} and Gl{\"o}{\ss}, Andreas},
  journal={Journal of Chemical Information and Modeling},
  volume={59},
  number={11},
  pages={4806--4813},
  year={2019},
  publisher={ACS Publications}
}

@article{mu2007performance,
  title={Performance of COSMO-RS with sigma profiles from different model chemistries},
  author={Mu, Tiancheng and Rarey, J{\"u}rgen and Gmehling, J{\"u}rgen},
  journal={Industrial \& engineering chemistry research},
  volume={46},
  number={20},
  pages={6612--6629},
  year={2007},
  publisher={ACS Publications}
}

@article{mulliken1955electronic,
  title={Electronic population analysis on LCAO--MO molecular wave functions. I},
  author={Mulliken, Robert S},
  journal={The Journal of chemical physics},
  volume={23},
  number={10},
  pages={1833--1840},
  year={1955},
  publisher={American Institute of Physics}
}

@article{person1974dipole,
  title={Dipole moment derivatives and infrared intensities. I. Polar tensors},
  author={Person, Willis B and Newton, James H},
  journal={The Journal of Chemical Physics},
  volume={61},
  number={3},
  pages={1040--1049},
  year={1974},
  publisher={American Institute of Physics}
}

@article{abranches2024stochastic,
  title={Stochastic machine learning via sigma profiles to build a digital chemical space},
  author={Abranches, Dinis O and Maginn, Edward J and Col{\'o}n, Yamil J},
  journal={Proceedings of the National Academy of Sciences},
  volume={121},
  number={31},
  pages={e2404676121},
  year={2024},
  publisher={National Academy of Sciences}
}

@article{schutt2017schnet,
  title={Schnet: A continuous-filter convolutional neural network for modeling quantum interactions},
  author={Sch{\"u}tt, Kristof and Kindermans, Pieter-Jan and Sauceda Felix, Huziel Enoc and Chmiela, Stefan and Tkatchenko, Alexandre and M{\"u}ller, Klaus-Robert},
  journal={Advances in neural information processing systems},
  volume={30},
  year={2017}
}

@article{duvenaud2015convolutional,
  title={Convolutional networks on graphs for learning molecular fingerprints},
  author={Duvenaud, David K and Maclaurin, Dougal and Iparraguirre, Jorge and Bombarell, Rafael and Hirzel, Timothy and Aspuru-Guzik, Al{\'a}n and Adams, Ryan P},
  journal={Advances in neural information processing systems},
  volume={28},
  year={2015}
}

@article{tayyebi2023prediction,
  title={Prediction of organic compound aqueous solubility using machine learning: a comparison study of descriptor-based and fingerprints-based models},
  author={Tayyebi, Arash and Alshami, Ali S and Rabiei, Zeinab and Yu, Xue and Ismail, Nadhem and Talukder, Musabbir Jahan and Power, Jason},
  journal={Journal of Cheminformatics},
  volume={15},
  number={1},
  pages={99},
  year={2023},
  publisher={Springer}
}

@article{xie2020improvement,
  title={Improvement of prediction performance with conjoint molecular fingerprint in deep learning},
  author={Xie, Liangxu and Xu, Lei and Kong, Ren and Chang, Shan and Xu, Xiaojun},
  journal={Frontiers in pharmacology},
  volume={11},
  pages={606668},
  year={2020},
  publisher={Frontiers Media SA}
}

@article{tsai2023improved,
  title={Improved vapor pressure prediction from PR+ COSMOSAC EOS using normal boiling temperature},
  author={Tsai, Chang-Che and Lin, Shiang-Tai},
  journal={AIChE Journal},
  volume={69},
  number={3},
  pages={e17997},
  year={2023},
  publisher={Wiley Online Library}
}

@article{dong2018united,
  title={A united chemical thermodynamic model: COSMO-UNIFAC},
  author={Dong, Yichun and Zhu, Ruisong and Guo, Yanyan and Lei, Zhigang},
  journal={Industrial \& Engineering Chemistry Research},
  volume={57},
  number={46},
  pages={15954--15958},
  year={2018},
  publisher={ACS Publications}
}

@article{dong2020cosmo,
  title={COSMO-UNIFAC model for ionic liquids},
  author={Dong, Yichun and Huang, Shuai and Guo, Yanyan and Lei, Zhigang},
  journal={AIChE Journal},
  volume={66},
  number={1},
  pages={e16787},
  year={2020},
  publisher={Wiley Online Library}
}

@article{hsieh2010improvements,
  title={Improvements of COSMO-SAC for vapor--liquid and liquid--liquid equilibrium predictions},
  author={Hsieh, Chieh-Ming and Sandler, Stanley I and Lin, Shiang-Tai},
  journal={Fluid Phase Equilibria},
  volume={297},
  number={1},
  pages={90--97},
  year={2010},
  publisher={Elsevier}
}

@article{paulechka2015reparameterization,
  title={Reparameterization of COSMO-SAC for phase equilibrium properties based on critically evaluated data},
  author={Paulechka, Eugene and Diky, Vladimir and Kazakov, Andrei and Kroenlein, Kenneth and Frenkel, Michael},
  journal={Journal of Chemical \& Engineering Data},
  volume={60},
  number={12},
  pages={3554--3561},
  year={2015},
  publisher={ACS Publications}
}

@article{mu2007group,
  title={Group contribution prediction of surface charge density profiles for COSMO-RS (Ol)},
  author={Mu, Tiancheng and Rarey, J{\"u}rgen and Gmehling, J{\"u}rgen},
  journal={AIChE journal},
  volume={53},
  number={12},
  pages={3231--3240},
  year={2007},
  publisher={Wiley Online Library}
}

@article{hsieh2012first,
  title={First-principles prediction of phase equilibria using the PR+ COSMOSAC equation of state},
  author={Hsieh, Chieh-Ming and Lin, Shiang-Tai},
  journal={Asia-Pacific Journal of Chemical Engineering},
  volume={7},
  pages={S1--S10},
  year={2012},
  publisher={Wiley Online Library}
}

@article{mahmoudabadi2021predictive,
  title={A predictive PC-SAFT EOS based on COSMO for pharmaceutical compounds},
  author={Mahmoudabadi, Samane Zarei and Pazuki, Gholamreza},
  journal={Scientific Reports},
  volume={11},
  number={1},
  pages={6405},
  year={2021},
  publisher={Nature Publishing Group UK London}
}

@article{niederquell2018new,
  title={New prediction methods for solubility parameters based on molecular sigma profiles using pharmaceutical materials},
  author={Niederquell, Andreas and Wyttenbach, Nicole and Kuentz, Martin},
  journal={International journal of pharmaceutics},
  volume={546},
  number={1-2},
  pages={137--144},
  year={2018},
  publisher={Elsevier}
}

@article{lee2007prediction,
  title={Prediction of mixture vapor--liquid equilibrium from the combined use of Peng--Robinson equation of state and COSMO-SAC activity coefficient model through the Wong--Sandler mixing rule},
  author={Lee, Ming-Tsung and Lin, Shiang-Tai},
  journal={Fluid Phase Equilibria},
  volume={254},
  number={1-2},
  pages={28--34},
  year={2007},
  publisher={Elsevier}
}

@article{huber2021ms2deepscore,
  title={MS2DeepScore: a novel deep learning similarity measure to compare tandem mass spectra},
  author={Huber, Florian and van der Burg, Sven and van der Hooft, Justin JJ and Ridder, Lars},
  journal={Journal of cheminformatics},
  volume={13},
  number={1},
  pages={84},
  year={2021},
  publisher={Springer}
}

@article{blaschke2020memory,
  title={Memory-assisted reinforcement learning for diverse molecular de novo design},
  author={Blaschke, Thomas and Engkvist, Ola and Bajorath, J{\"u}rgen and Chen, Hongming},
  journal={Journal of cheminformatics},
  volume={12},
  number={1},
  pages={68},
  year={2020},
  publisher={Springer}
}

@article{jirasek2021perspective,
  title={Perspective: machine learning of thermophysical properties},
  author={Jirasek, Fabian and Hasse, Hans},
  journal={Fluid Phase Equilibria},
  volume={549},
  pages={113206},
  year={2021},
  publisher={Elsevier}
}

@article{jirasek2023combining,
  title={Combining machine learning with physical knowledge in thermodynamic modeling of fluid mixtures},
  author={Jirasek, Fabian and Hasse, Hans},
  journal={Annual review of chemical and biomolecular engineering},
  volume={14},
  number={1},
  pages={31--51},
  year={2023},
  publisher={Annual Reviews}
}

@article{hayer2025advancing,
  title={Advancing thermodynamic group-contribution methods by machine learning: UNIFAC 2.0},
  author={Hayer, Nicolas and Wendel, Thorsten and Mandt, Stephan and Hasse, Hans and Jirasek, Fabian},
  journal={Chemical Engineering Journal},
  volume={504},
  pages={158667},
  year={2025},
  publisher={Elsevier}
}

@article{hayer2025modified,
  title={Modified UNIFAC 2.0-A Group-Contribution Method Completed with Machine Learning},
  author={Hayer, Nicolas and Hasse, Hans and Jirasek, Fabian},
  journal={Industrial \& Engineering Chemistry Research},
  volume={64},
  number={20},
  pages={10304--10313},
  year={2025},
  publisher={ACS Publications}
}

@article{hoffmann2025machine,
  title={A machine-learned expression for the excess Gibbs energy},
  author={Hoffmann, Marco and Specht, Thomas and Göttl, Quirin and Burger, Jakob and Mandt, Stephan and Hasse, Hans and Jirasek, Fabian},
  journal={arXiv preprint arXiv:2509.06484},
  year={2025}
}

@article{soares2025recent,
  title={Recent Improvements to the NWChem COSMO Module},
  author={Soares, Rafael de P and Mej{\'\i}a-Rodriguez, Daniel and Apr{\`a}, Edoardo},
  journal={Journal of Chemical Theory and Computation},
  volume={21},
  number={22},
  pages={11573--11584},
  year={2025},
  publisher={ACS Publications}
}

@article{apra2020nwchem,
  title={NWChem: Past, present, and future},
  author={Apra, Edoardo and Bylaska, Eric J and De Jong, Wibe A and Govind, Niranjan and Kowalski, Karol and Straatsma, Tjerk P and Valiev, Marat and van Dam, Hubertus JJ and Alexeev, Yuri and Anchell, James and others},
  journal={The Journal of chemical physics},
  volume={152},
  number={18},
  year={2020},
  publisher={AIP Publishing}
}

@article{gerlach2022open,
  title={An open source COSMO-RS implementation and parameterization supporting the efficient implementation of multiple segment descriptors},
  author={Gerlach, Thomas and M{\"u}ller, Simon and de Castilla, Andr{\'e}s Gonz{\'a}lez and Smirnova, Irina},
  journal={Fluid phase equilibria},
  volume={560},
  pages={113472},
  year={2022},
  publisher={Elsevier}
}
\clearpage
\begin{figure}[H]
    \centering
    \includegraphics[width=\textwidth]{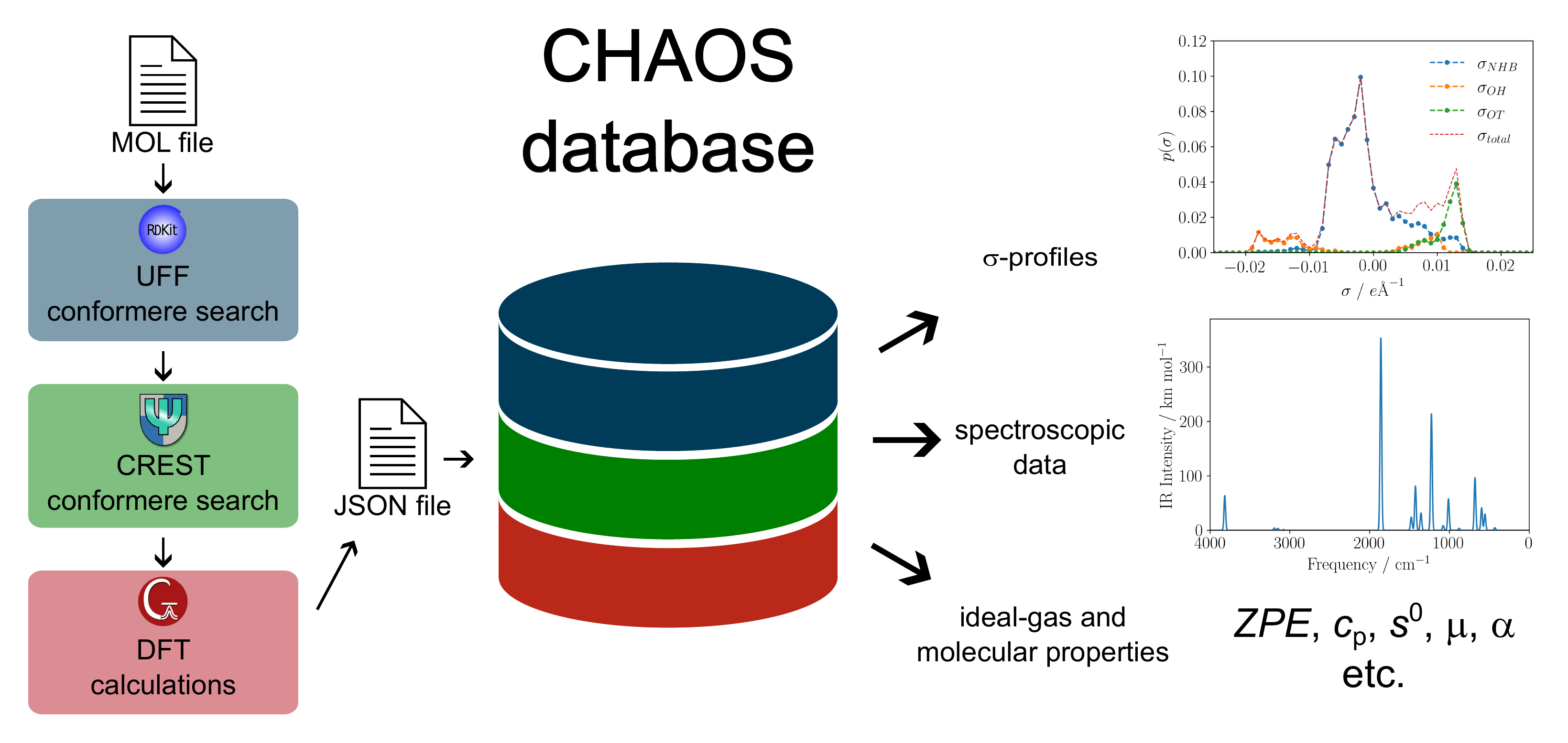}
    \caption{TOC figure.}
    \label{fig:TOC}
\end{figure}
\end{document}